# Exploring Low Internal Reorganization Energies for Silicene Nanoclusters

*Ricardo Pablo-Pedro[1], Hector Lopez-Rios[2], Jose-L Mendoza-Cortes[3,4], Jing Kong[5], Serguei Fomine[2], Troy Van Voorhis[1,*], and Mildred S. Dresselhaus[5,6,¶]*

[1] Department of Chemistry, Massachusetts Institute of Technology, 77 Massachusetts Avenue, Cambridge, MA 02139, United States

[2] Instituto de Investigaciones en Materiales, Universidad Nacional Autónoma de México, Apartado Postal 70-360, CU, Coyoacán, Ciudad de México 04510, México

[3] Department of Chemical & Biomedical Engineering, Florida A&M University and Florida State University, Joint College of Engineering, Tallahassee FL, 32310, USA

[4] Scientific Computing Department, Materials Science and Engineering Program, High Performance Material Institute, Condensed Matter Theory, National High Magnetic Field Laboratory, Florida State University, Tallahassee FL, 32310, USA

[5] Department of Electrical Engineering and Computer Science, Massachusetts Institute of Technology, Cambridge, MA 02139, United States

[6] Department of Physics and Department of Electrical Engineering and Computer Science, Massachusetts Institute of Technology, Cambridge, MA 02139, United States

### Abstract

High-performance materials rely on small reorganization energies to facilitate both charge separation and charge transport. Here, we performed DFT calculations to predict small reorganization energies of rectangular silicene nanoclusters with hydrogen-passivated edges denoted by H-SiNC. We observe that across all geometries, H-SiNCs feature large electron affinities and highly stabilized anionic states, indicating their potential as n-type materials. Our findings suggest that fine-tuning the size of H-SiNCs along the "zigzag" and "armchair" directions may permit the design of novel n-type electronic materials and spinctronics devices that incorporate both high electron affinities and very low internal reorganization energies.

# Introduction

Graphene has attracted increasing interest since its discovery for its potential as a future elementary unit in modern microelectronics, such as field-effect transistors [1], photovoltaic cells [2], advanced gas sensors [3], and battery energy storage [4]. Additionally, the isolation of graphene has inspired the new world of two-dimensional (2D) materials, where some recent members include black phosphorus [5], transition metal dichalcogenides (TMDs) [6], and silicene [7]. Among these 2D materials, silicene has been predicted to possess most of the remarkable electronic properties of graphene, such as Dirac cones, carrier mobility, and high Fermi velocity [7]. Moreover, the major advantage of investigating silicene is that it can be easily incorporated into the present silicon-based microelectronics industry [8,9]. In this direction, considerable improvements have been achieved in the use of silicene as field-effect transistors at room temperature [10] and in spintronics applications [11].

Interestingly, the electronic properties of silicene change in a notrivial manner when going to lower dimensions. Silicene nanoribbons, for instance, can be either metals or semiconductors depending on width [9]. For the case of silicene nanoclusters, the electronic structure is expected to be different from pristine silicene and nanoribbons because there are additional degrees of freedom from the edge atoms [8]. In addition, silicene nanoclusters have the advantage that they all possess fine band gaps because of quantum confinement effect that is desirable for their applications. The fabrication of these silicene nanostructures could be realized using etching techniques that have been already used in cutting graphene sheets into nanostructures with desired shape and size. This experimental progress motivates our study of electronic structure of silicene nanoclusters with rectangular shape [12,13] .

Carrier transport is the central property for applications of silicene in field-effect transistors, thermoelectric devices, and spintronics. Therefore, having high carrier mobility is very important for developing new silicene nanoelectronic devices. The transport mechanism is generally understood with respect to the limiting cases of hopping (polaron) transport and band-like transport corresponding to the extreme localization or delocalization of the charge carriers. The band-like regime is generally observed only at low temperatures in highly ordered samples (i.e. the mobility $\mu$ decreases with temperature as $\mu \propto T^{-n}$) up to room temperature [14] . The hopping (polaron) mechanism consists when the carrier is localized on one molecule through the formation of a self-trapped state (a polaron) and transport occurs through a thermally activated hopping mechanism (strong electron-phonon coupling), which is observed commonly in small molecules and molecular semiconductors at high temperatures [15]. For instance, the electron-phonon interaction of silicene is considerable, as the larger Si-Si interatomic distance of silicene, compared to graphene, weakens the pi-pi overlaps and results in a low-buckled structure with sp3-like hybridization [7]. Then, because of the size of our silicene nanoclusters and large geometry relaxations of the silicene nanoclusters, we consider that they could be described by the hopping mechanism in which the mobility ($\mu$) of the charge carriers (electrons or holes) is directly proportional to the transfer rate ($k$) of charge carriers described by the Einstein relation [16]

$$\mu = \frac{eA^2}{k_B T} k \quad (1)$$

Where $k_B$, T, e and A are the Boltzmann constant, temperature, the electron charge, and the hopping transport distance. According to the Marcus-Huss semiclassical model for the inter-chain charge transfer, the charge transfer rate is given by [16]

$$k = \frac{2\pi}{h} \left(\frac{\pi}{\lambda k_B T}\right)^{-1/2} H_{ab}^2 e^{-\left(\frac{\lambda}{4k_B T}\right)^2} \quad (2)$$

Here h, $\lambda$, and $H_{ab}$ are the Plank constant, the reorganization energy and the electronic coupling matrix element between the donor and acceptor, respectively. According to Eq. 2, the charge transfer is determined by $H_{ab}$ and $\lambda$. However, $H_{ab}$ starts to saturate for oligomers with more than 5 monomeric units indicating that the exponential contribution in Eq. 2 keeps increasing and for more than 7 monomeric units it dominates the behavior of the charge transfer rate [17]. Therefore, the key parameter that governs the behavior of the charge transfer rate is the reorganization energy which decreases with increasing length of the system, as will be discussed below. At some point the transfer coupling is expected to saturate once it depends on the conjugation length of polymers [16]. Then, for large conjugated polymers and organic molecular semiconductors, the exponential nature of Eq. 2 dominates and in this way, the reorganization energy is the most important parameter to be studied to estimate the charge carrier transport [16,17]. The reorganization energy, $\lambda$, corresponds to the energy of structural change when going from neutral-state geometry to charged-state geometry and vice versa. In terms of the electron-phonon mechanism, $\lambda$ is a measure of the strength of the hole-phonon or electron-phonon interaction and produces a relation between the geometry, the electronic structure, and the transport properties of the material [19]. The reorganization energy can be divided in terms of the transport of holes ($\lambda_+$) and electrons ($\lambda_-$) in order to determine whether a material may have a greater mobility of electrons than holes or of holes than electrons. It is known that most organic semiconductors have hole reorganization ($\lambda_+$) energies greater than 0.1 eV, however several hole-transporting (*p*-type) organic semiconductors have been reported with hole reorganization energies of less than 0.1 eV, for example, pentacene (0.097 eV) [20] and circum(oligo)acenes (0.057 to 0.127 eV) [21]. On the other hand, studies of electron-transporting (*n*-type) semiconductors have been limited in the last decade because of the instability in air of radical anionic semiconductors and the high injection barrier of electrons. One of the few examples include fullerene ($C_{60}$), which has been found to be an excellent *n*-type acceptor with a small electron reorganization energy of $\lambda_- = 0.060$ eV [22]. This allows fast photoinduced charge separation and slow charge recombination resulting in the formation of a long-lived charge-separated state with high quantum yield [23]. The small reorganization energy of fullerene can also be ascribed to its extended π-conjugation and its rigid molecular structure. In general, the hole ($\lambda_+$) and electron ($\lambda_-$) reorganization energy decreases as the size of the π conjugated system is increased. For instance, anthracene $\lambda_+ = 0.137$ eV and $\lambda_- = 0.196$ eV; tetracene, 0.113 eV and 0.160 eV; and pentacene, 0.097 eV and 0.132 eV, [24] respectively.

In anisotropic two-dimensional (2D) semiconductors such as TMDs [25,26], black phosphorus [27,28], and silicene [8]; the electrons and phonons behave differently in different in-plane directions, e.g., armchair and zigzag directions. This leads to angle-dependent mechanical, optical, and electrical responses. Therefore, we hypothesized that the reorganization energies in anisotropic materials should exhibit different behaviors along different directions. These unique properties may permit the design of novel electronics and optoelectronics with anisotropic crystalline orientation.

In this work, we study reorganization energies of H-SiNCs along different directions by using density functional theory (DFT). The values of the reorganization energies for H-SiNCs are obtained by varying the width of the zigzag and the armchair directions. To further evaluate these H-SiNCs, we computed the adiabatic electron affinities (EAs) and the ionization potential (IP) energies. To the best of our knowledge, this is the first theoretical work exploring the reorganization energies for different directions of rectangular silicene nanoclusters.

# Computational Details

We have carried out geometrical optimizations of rectangular silicene nanoclusters utilizing the three-parameter Becke and Lee-Yang-Parr functional (B3LYP) in conjunction with Dunning's correlation-consistent cc-pVDZ basis set for all atoms. For all molecules we used restricted DFT (RB3LYP) or unrestricted DFT (UB3LYP) when specified. Grimme's D3 dispersion correction [29] was also used in all DFT calculations to account for the dispersion interactions. Thus, the methods are named RB3LYP-D3 and UB3LYP-D3; we indicate in the text if only one method is used. The need for UB3LYP-D3 is due to the singlet-triplet instability detected in the model's optimized closed-shell singlet state wave function, also necessitating the consideration of two different values of multiplicities, singlet and triplet states. Here, a singlet state is an electronic state such that all electron spins are paired (anti-ferromagnetic state) while in a triplet state the electron that is promoted has the same spin orientation (parallel) to the other unpaired electron (ferromagnetic state). Singlet and triplet spin values are derived using the equation for angular momenta multiplicity, 2S+1, where S is the total spin angular momentum (sum of all electron spins in the molecule). The closed-shell singlet systems that presented triplet instability were re-optimized using a broken-symmetry unrestricted (UBS) method [30]. Additionally, frequency calculations for selected H-SiNCs were performed to ensure the absence of any vibrational instabilities in the ground state structures. None of the systems analyzed have negative frequencies, which means that they are thermodynamically stable. The calculations were performed using the code TURBOMOLE V7.0 [31]. From here on, we will denote UB3LYP-D3 method as B3LYP for simplicity.

We also ran complete active-space self-consistent field (CASSCF) calculations of the charged molecules of the corresponding electronic multiplicity also using the B3LYP functional to explore their multi-configurational character. This is due to correct and compare, if needed, for the multi-configurational character of the neutral H-SiNCs reported in a previous study [8] at the B3LYP level of theory. The chosen active spaces consisted of 9 electrons and 10 orbitals, and 11 electrons and 10 orbitals, for cation and anion radicals, respectively. This active space was the largest practical active space possible. In addition, we visualized the corresponding orbitals to ensure that they involve all atoms of studied systems, and we inspected the density matrix generated in each calculation to ensure that all active orbitals have population numbers between 0.1 and 1.9; confirmation of a correctly selected active space. The 6-31G(d) basis set was assigned to all atoms. These calculations were carried out with Gaussian 09 rD.01. code [32].

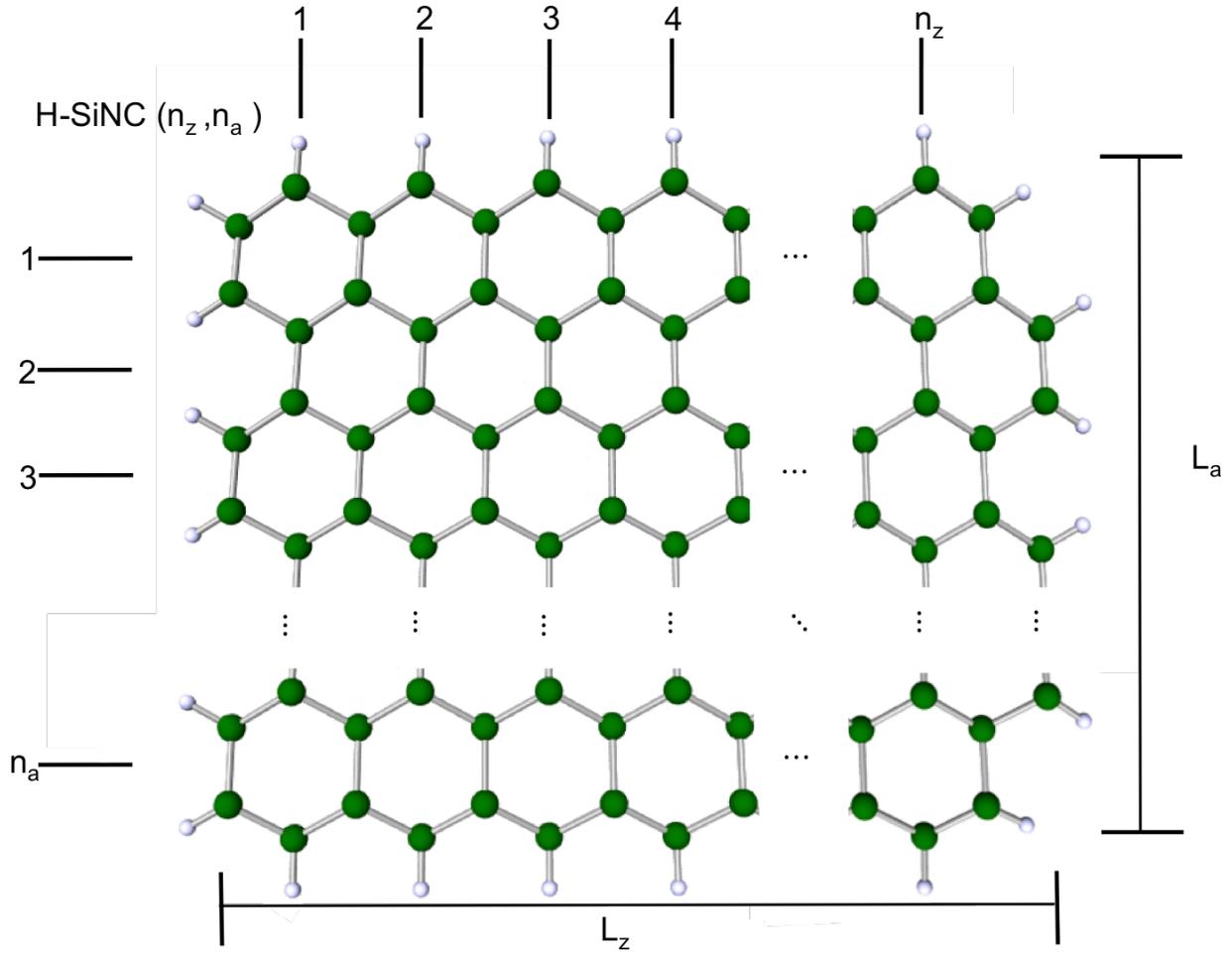

**Figure 1**. Top-down view of a lattice structure of silicene, where the edges are passivated with hydrogen. The $n_z$ and $n_a$ are the number of fused benzene rings in columns and rows.

We study rectangular silicene nanoclusters following the pattern shown in Figure 1. All edge dangling bonds were passivated with hydrogen. To specify the size of each H-SiNC, we use $n_z$ and $n_a$, which correspond to the number of hexagonal units along the zigzag and armchair edges. Here we define a characteristic area scale by $N = n_z \times n_a$, which corresponds to the total of fused rings in the silicene nanocluster. The characteristic dimensions of each H-SiNC are denoted here by $L_a$ (armchair edge) and $L_z$ (zigzag edge), being linear functions of $n_a$ and $n_z$, and the average Si-Si bond length. We vary $n_a$ from 1 to 9 ( which means that $L_a$ varies from 0.439 nm for $n_a$=1 to 3.112 nm for N=9, approximately ) and for $n_z$ from 1 to 7 ($L_z$ varying from 0.381 nm for $n_z$=1 to 2.882 nm for $n_z$=7).

The geometry of every silicene nanocluster was first optimized for the neutral molecule at the B3LYP level, $E_0(q_0)$. The energy was then calculated for the negatively charged molecule (adding an electron to the neutral molecule) in the optimized geometry of the neutral molecule, $E_-(q_0)$. The geometry of the negatively charged molecule was subsequently optimized to obtain $E_-(q_-)$. Their energy difference, $E_-(q_0) - E_-(q_-)$, is equal to the reorganization energy $\lambda_1$, see Fig. 2. We now remove an electron from the optimized negatively charged molecule, and calculate the energy without relaxing the geometry to obtain the $E_0(q_-)$ state. The energy difference between the $E_0(q_-)$ state and the optimized neutral molecule $E_0(q_0)$ is the reorganization energy $\lambda_2$, as shown in Fig. 2. Then, the total electron reorganization energy for the process is equal to [33]

$$\lambda_- = \lambda_2 + \lambda_1 = [E_0(q_-) - E_0(q_0)] + [E_-(q_0) - E_-(q_-)] \qquad (3)$$

where E is energy and q is the geometry. The subscripts 0 and – represent the neutral and anionic states, respectively. The same procedure was carried out for the cation state. Thus, the hole reorganization energy is given by [33]

$$\lambda_+ = \lambda_4 + \lambda_3 = [E_0(q_+) - E_0(q_0)] + [E_+(q_0) - E_+(q_+)] \qquad (4)$$

Here $E$ is energy and $q$ is the geometry as in Eq. 3. The subscripts 0 and + represent the neutral and cationic states, respectively. Moreover, we used total-energy differences between the DFT-B3LYP calculations performed for the neutral and charged systems to evaluate: (i) the adiabatic ionization potentials (IPs) and electron affinities (EAs); (ii) the highest occupied molecular orbital (HOMO) and lowest unoccupied molecular orbital (LUMO) energies for the neutral systems. The vertical transitions involved in Equations 3 and 4 are shown in Figure 2.

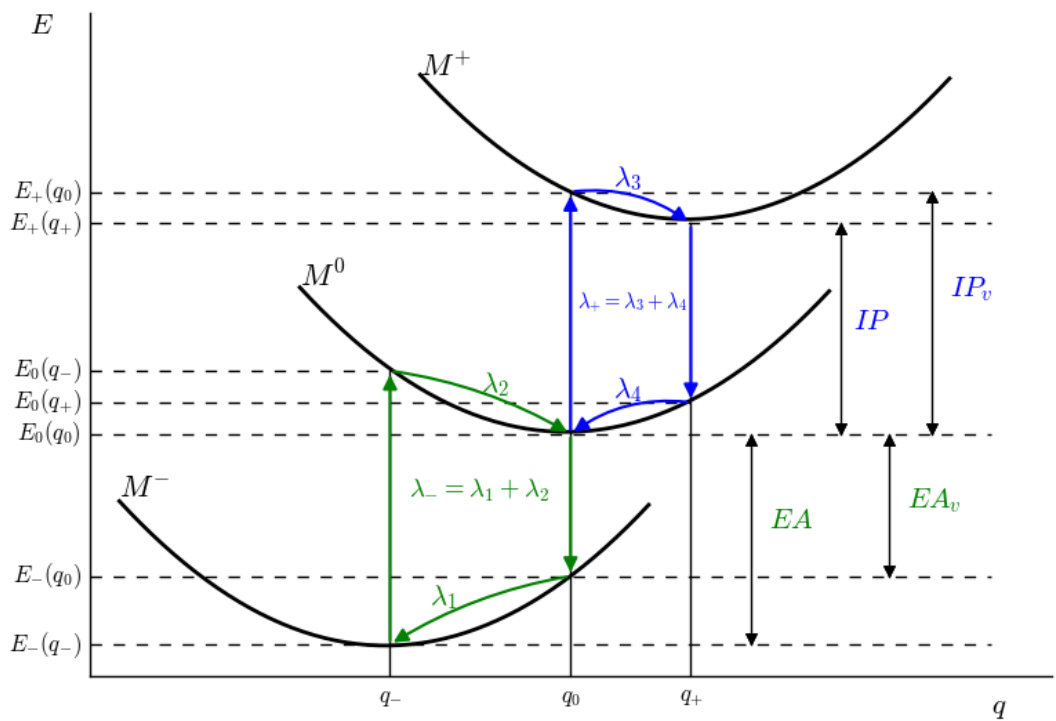

**Figure 2**. Potential energy surfaces sketch for the neutral ($M^0$), negatively ($M^-$) and positively ($M^+$) charged structures. E is the energy, q is the geometry, and the subscripts 0, -, and + denote neutral, anionic and cationic states, respectively. Here IP is the ionization potential, EA is the electron affinity, and the subscript $v$ denotes vertical energy transitions.

# Results and discussion

We summarize the optimized bond lengths of ground, cationic, and anionic states for selected H-SiNC($n_z, n_a$) in Table 1 at the B3LYP/cc-pVDZ level of theory. Two multiplicities were used according to previous results, i.e., single (S=0) and triplet (S=1). The bond length increases (decreases) in going from the neutral to the negatively (positively) charged structure show a consistent trend among the series. Overall, small molecules show the largest geometry relaxations, e.g., H-SiNC(2,1) changes its Si-Si bond lengths approximately 0.025 Å. This value is reduced to 0.005 Å and 0.002 Å in H- SiNC(7,1) and H-SiNC(5,4), respectively. However, the geometry distortions in the buckled structure of the silicene nanoclusters are larger than in the Si-Si bond lengths with averaged differences in the range of ±0.03 Å, indicating that even for a large system, e.g., H-SiNC(5,4), the neutral molecule will suffer large height modifications in its charged species, as shown in Fig. 3. Compared to the neutral geometry, the cationic geometry shrinks via the shortening of both bonds and heights, and the anionic geometry expands via lengthening of the bonds and heights. These geometry distortions (i.e. bond lengths and heights) of the H-SiNCs could be a key factor to obtain small reorganization energies. For instance, fullerenes have an electron reorganization energy ($\lambda_-$) that is particularly low because they are extremely rigid [22]. In terms of π-conjugation, larger silicene nanoclusters possess an extensive π conjugation that leads to greater delocalized charge distribution. Thus, local structural adjustment (i.e., bond length and height) for electron transfer ($\lambda_-$) is less severe in large silicene systems than smaller systems; see the Si-Si bond lengths and heights ($\Delta_a$) in Table 1. In general, for the H-SiNCs, the largest structural changes are observed in the average height of the anions.

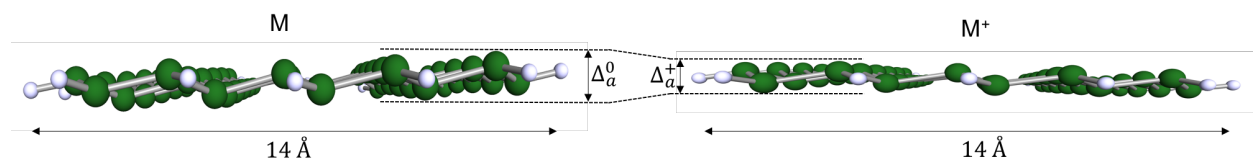

**Figure 3**. Neutral state (left) of H-SiNC(5,4) relaxes into a new structure (right) when one

electron is removed. As a result, the silicene nanocluster contracts along the vertical axis. The displacement is magnified 2 times in order to make this distortion clear.

We summarize the calculated hole ($\lambda_+$) and electron ($\lambda_-$) reorganization energies, the adiabatic values of electron affinity (EA), and the ionization potential (IP) for the rectangular silicene nanoclusters in Table 2 at the B3LYP level of study. It has been demonstrated that the B3LYP functional used for the reorganization energy calculations best reproduces experimental data for conjugated organic systems [34,35]. In addition, the B3LYP functional was found to reproduce reliable EAs for 14 atoms and 96 molecules [36].

**Table 1**. The average Si-Si bond lengths ($d_a$) and the average height ($\Delta_a$) of selected neutral, cationic and anionic silicene structures at the B3LYP level of calculation. Here the subscripts 0, +, and − represent the neutral, cationic, and anionic states, respectively. Here S represents the initial triplet and singlet states.

| H-SiNC($n_z, n_a$) | S | $d_a^0$[A] | $d_a^+$[A] | $d_a^-$[A] | $\Delta_a^0$[A] | $\Delta_a^+$[A] | $\Delta_a^-$[A] |
|---|---|---|---|---|---|---|---|
| (2,1) | 0 | 2.254 | 2.251 | 2.279 | 0.459 | 0.396 | 0.572 |
| (7,1) | 0 | 2.264 | 2.260 | 2.269 | 0.441 | 0.420 | 0.479 |
| (5,4) | 0 | 2.271 | 2.269 | 2.273 | 0.480 | 0.467 | 0.489 |
| (4,7) | 0 | 2.272 | 2.270 | 2.273 | 0.482 | 0.473 | 0.487 |
| (4,9) | 0 | 2.272 | 2.271 | 2.273 | 0.480 | 0.475 | 0.484 |
| (4,9) | 1 | 2.272 | 2.271 | 2.273 | 0.480 | 0.474 | 0.488 |

The adiabatic IPs and EAs presented in Table 2 were computed from the difference in the total energy between the optimized neutral state and the corresponding optimized cation or anion state, i.e., IP = $E_0(q_0)$ - $E_+(q_+)$ and EA = $E_0(q_0)$ - $E_-(q_-)$ [37]. For all the studied H-SiNCs, the cationic-state potential energy was higher than the neutral-state potential energy giving a positive

IP. Moreover, it is observed that the IPs of the H-SiNCs drop with the number of fused rings, N, which is in agreement with other conjugated systems. In the case of the EAs, all H-SiNCS will bind an electron with a positive EA, which also increases along with the number of benzoic rings. For *n*-type organic semiconductors, theadiabatic EA is an important property that determines organic field-effect transistor device performance such as durability. Among the studied systems, a few H-SiNCs exhibit quite large adiabatic EAs, exceeding the threshold of 2.80 eV used for classifying *n*-type materials [38,39]. Hence, H-SiNCs are expected to be stable electron transport materials [40].

**Table 2**. Hole ($\lambda_+$) and electron ($\lambda_-$) reorganization energies, along with their adiabatic ionization potentials (IPs) and electron affinities (EAs) calculated at the B3LYP level of theory. Here $S_{initial}$ represents the initial triplet and singlet states.

| H-SiNC($n_z, n_a$) | N | $S_{initial}$ | $\lambda_-[eV]$ | EAs[eV] | $\lambda_+[eV]$ | IPs[eV] |
|---|---|---|---|---|---|---|
| (1,1) | 1 | 0 | 0.775 | 1.39 | 0.303 | 7.12 |
| (2,1) | 2 | 0 | 0.297 | 1.93 | 0.182 | 6.47 |
| (3,1) | 3 | 0 | 0.237 | 2.39 | 0.148 | 6.06 |
| (4,1) | 4 | 0 | 0.121 | 2.68 | 0.117 | 5.80 |
| (5,1) | 5 | 0 | 0.088 | 2.85 | 0.099 | 5.66 |
| (6,1) | 6 | 0 | 0.078 | 2.94 | 0.094 | 5.59 |
| (7,1) | 7 | 0 | 0.077 | 2.99 | 0.093 | 5.56 |
| (2,2) | 4 | 0 | 0.315 | 2.38 | 0.144 | 6.06 |
| (3,2) | 6 | 0 | 0.233 | 2.73 | 0.117 | 5.75 |
| (4,2) | 8 | 0 | 0.136 | 2.95 | 0.101 | 5.56 |
| (5,2) | 10 | 0 | 0.114 | 3.05 | 0.100 | 5.48 |
| (6,2) | 12 | 0 | 0.134 | 3.10 | 0.105 | 5.46 |
| (7,2) | 14 | 0 | 0.159 | 3.13 | 0.099 | 5.47 |
| (3,3) | 9 | 0 | 0.142 | 2.91 | 0.085 | 5.57 |
| (4,3) | 12 | 0 | 0.091 | 3.09 | 0.077 | 5.42 |
| (5,3) | 15 | 0 | 0.143 | 3.15 | 0.086 | 5.40 |
| (6,3) | 18 | 0 | 0.190 | 3.19 | 0.133 | 5.40 |
| (7,3) | 21 | 0 | 0.205 | 3.21 | 0.090 | 5.38 |
| (4,4) | 16 | 0 | 0.081 | 3.19 | 0.072 | 5.33 |
| (5,4) | 20 | 0 | 0.220 | 3.22 | 0.083 | 5.34 |
| (6,4) | 24 | 0 | 0.188 | 3.24 | 0.138 | 5.34 |
| (7,4) | 28 | 0 | 0.150 | 3.28 | 0.075 | 5.29 |
| (4,5) | 20 | 0 | 0.067 | 3.26 | 0.063 | 5.27 |
| (4,6) | 24 | 0 | 0.287 | 3.31 | 0.065 | 5.22 |
| (4,7) | 28 | 0 | 0.076 | 3.35 | 0.070 | 5.18 |
| (4,8) | 32 | 0 | 0.214 | 3.39 | 0.066 | 5.15 |
| (4,9) | 36 | 0 | 0.181 | 3.41 | 0.206 | 5.12 |
| (4,7) | 28 | 1 | 0.069 | 3.09 | 0.065 | 5.14 |
| (4,8) | 32 | 1 | 0.062 | 3.16 | 0.059 | 5.12 |
| (4,9) | 36 | 1 | 0.088 | 3.19 | 0.052 | 5.10 |

As seen in Table 2, almost all of the H-SiNCs have a larger $\lambda_-$ value than $\lambda_+$. Contextualizing our results, linear silicene nanoclusters could be considered as analogous molecules with oligoacenes. For instances, we compared some $\lambda_+$ values of linear H-SiNCs with some common polycylic aromatic hydrocarbons, i.e., $C_{10}H_8$ (N = 2), $C_{14}H_{10}$ (N = 3), $C_{18}H_{12}$ (N = 4), and $C_{22}H_{14}$ (N = 5) at B3LYP/6-31G** level of theory, which are 0.187, 0.137, 0.113, and 0.097 eV[12], respectively. These values are comparable with the corresponding silicene systems H-SiNC(2,1), 0.182 eV; H-SiNC(3,1), 0.148 eV; H-SiNC(4,1), 0.117 eV; and H-SiNC(5,1), 0.099 eV at the B3LYP/cc-pVDZ level of theory. For some H-SiNCs, their hole and electron reorganization energies are less than 0.1 eV such as H-SiNC(4,5) and H-SiNC(4,7). For systems with an initial triplet ground state, we can obtain even smaller hole and electron reorganization energies that can be compared to fullerene (0.06 eV). These low values suggest that the mobility along the armchair direction is greater than in the zigzag direction for these systems, with $n_a > n_z$. In addition, we can observe that the values of $\lambda_-$ and $\lambda_+$ for systems with an initial triplet ground (S=1) state are similar. This similarity in the electron and reorganization energies may suggest that these specific systems could act as ambipolars or n-acceptor materials merely considering the calculated electron and hole reorganization energies and adiabatic EA values. From these results it can be concluded that the initial ground state's spin state of the H-SiNCs has a profound effect on the reorganization energy as seen for H-SiNC(4,$n_a$), with $n_a$=7-9.

To examine the effect of the number of fused rings on the electron transfer, $\lambda_-$ is plotted for different H-SiNCs in Fig. 4 (a). Since we are comparing different orientations (i.e. armchair and zigzag directions), there is not a perfect relationship between electron reorganization energy and the number of fused rings, but there is a trend: electron reorganization generally decreased with increased N. At N = 24 we observed that for H-SiNC(6,4) and H-SiNC(4,6) structures, the change on the length in the armchair and zigzag direction significantly affects the electron and hole reorganization energies. The greatest decrease in $\lambda_-$ and $\lambda_+$ was obtained from molecules with initial triplet spin multiplicity. Thus, these low reorganization energies indicate an advantage for the H-SiNCs to transport electrons or holes when coupled to an external device that either donates or withdraws electrons such as a field effect transistor (FET) [10]. Another observation from Table 2 is that $\lambda_\pm$ follows the trend that larger clusters (more extended $\pi$ conjugation) have smaller $\lambda_\pm$ values. In the inset graph of Fig. 4, we illustrate the relationship between electron affinity (EA) and $\lambda_-$. It is observed that small H-SiNCs have low electron affinity and high electron reorganization energy. On the contrary, large H-SiNCs have high electron affinity and low electron reorganization energy, two important factors for good *n*-type materials [33]. In the case of linear or nearly linear H-SiNCs, Fig. 4 (b) shows that the reorganization energy for all H-SiNCs decreases as N increases. This relation is comparable with the $\lambda \sim N$ relation derived for linear oligomers [41], where N is the number of heterocyclic rings.

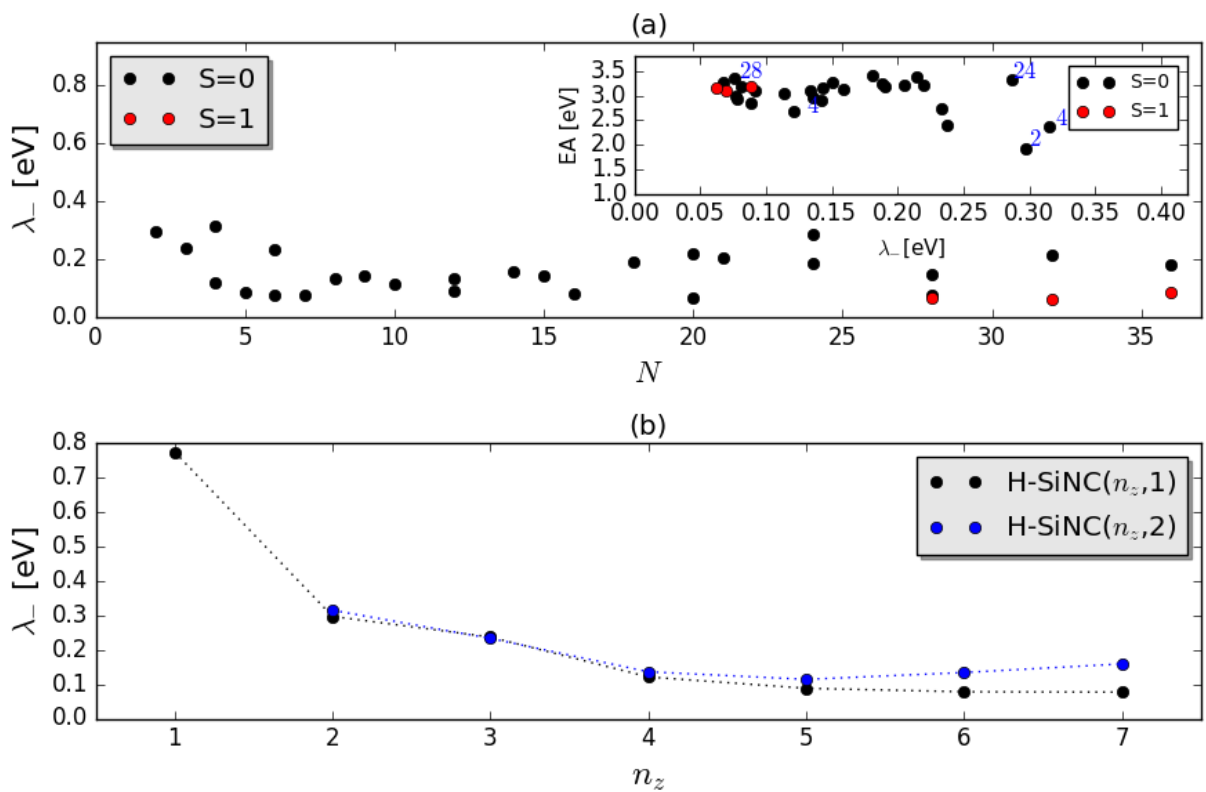

**Figure 4**. (a) Plot of electron reorganization energy ($\lambda_-$) versus number of fused rings N for H-SiNCs. Here S represents the initial spin for the triplet and singlet states for $S = 1$ and $S = 0$, respectively. Inset: Plot of electron affinity (EA) versus $\lambda_-$. The numbers in the graph indicate the number of fused rings (N) for selected H-SiNCs. (b) Plot of the electron reorganization energy versus number of fused rings along the zigzag direction for H-SiNCs with a singlet ground state.

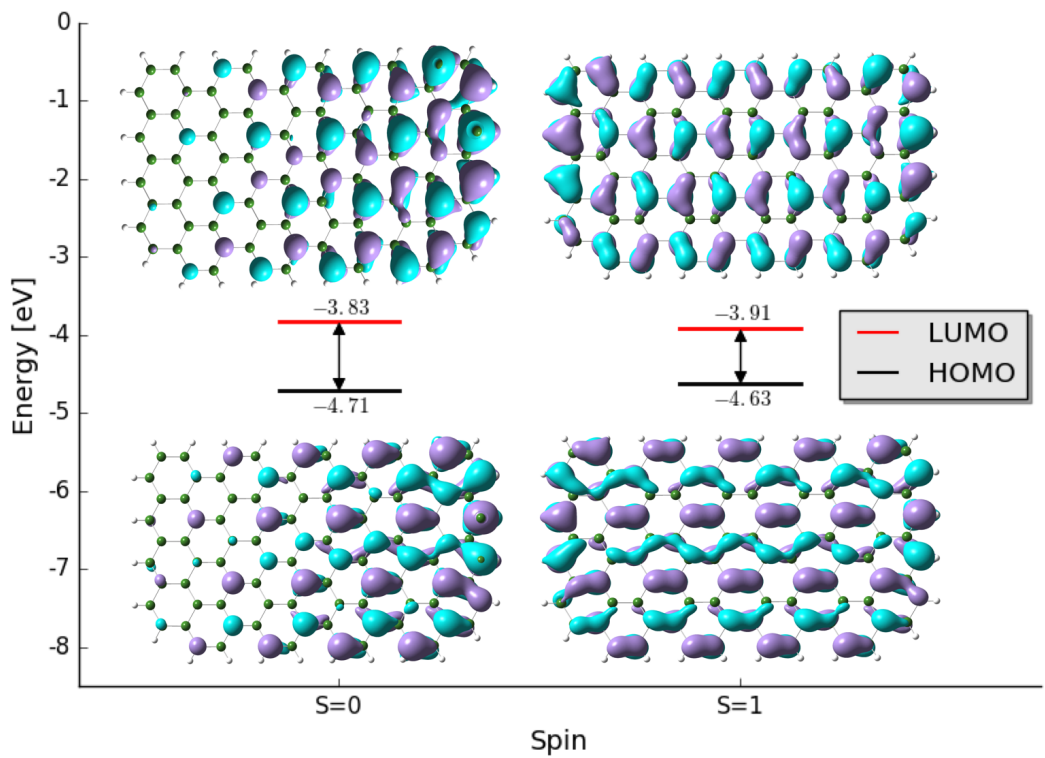

**Figure 5**. Frontier orbitals for neutral H-SiNC(4,9) with an initial singlet (S=0), and triplet (S=1) states at the B3LYP/cc-pVDZ level of theory. Molecular Orbitals were drawn with GaussView [32].

From our results, we observed that critical geometric changes of the neutral molecule can occur when an electron is added to the LUMO ($\lambda_-$) or when an electron it is removed from the HOMO ($\lambda_+$). The extra electron or hole in the neutral molecule will induce localized charge defects in the form of polarons (see the SI for more information). The polaron binding energy is from the deformations in lattice and molecular geometries that occur as the carrier localizes on a given site. This quantity is thus closely linked to the reorganization energy in electron-transfer theories.

This polaron energy arising from internal degrees of freedom can be obtained by expanding the site energy $\epsilon_n$ in powers of molecular normal-mode coordinates, which can be calculated from the frontier molecular orbitals (HOMO and LUMO for hole and electron transport, respectively). In addition, these geometric changes among charge transfer processes can correlate with the bonding character in the frontier orbitals [33]. This bonding character between two nearest atoms involving a molecular orbital is composed of bonding, antibonding, and nonbonding orbitals. For instance, adding an electron in the bonding orbital stabilizes the molecule because it is in between the atoms while adding an electron into the antibonding orbital will decrease the stability of the molecule. However, nonbonding only involves one atom, so the addition or removal of an electron does not change the energy or bond order (bond length alteration) of the molecule. Interestingly, it has been demonstrated that by having a high nonbonding character [33], we can obtain low reorganization energies because nonbonding induces much less bond length adjustment than bonding and anti-bonding upon charge transfer. Therefore, to examine whether structural adjustments with charge transfer correlates with the spatial distribution of frontier orbitals of the H-SiNCs to obtain low reorganization energies; we show in Fig. 5 that the percentage of nonbonding character is smaller in the HOMO than in the LUMO for H-SiNC(4,9) with an initial singlet state (see the supplementary for more information about bond length alternations). Therefore, this factor should contribute to a larger value of $\lambda_+$ than $\lambda_-$. Comparing the height of the cation and anion for H-SiNC(4,9) with $S_{initial}=0$, we observe that the height difference between the neutral and cation, $|\Delta_{C-N}| = 0.005$ (Table 1), is larger than that of $|\Delta_{A-N}| = 0.004$, as with the values of $\lambda_+$ relative to $\lambda_-$. In Figure 5, it is shown that when more extended delocalization is present for the initial triplet state molecule, the extended delocalization leads to the reduction of $\lambda_+$ and $\lambda_-$. For H-SiNC(4,9) with $S_{initial}=1$ ( see Fig. 5), it is noted that the values of $|\Delta_{A-N}| = 0.008$ is larger than that of $|\Delta_{C-N}| = 0.006$ ( Table 1), as with the values of $\lambda_-$ relative to $\lambda_+$. Therefore, this shows how specific structural modifications such as bond length alterations are originated upon the removal and addition of an electron to the silicene nanoclusters.

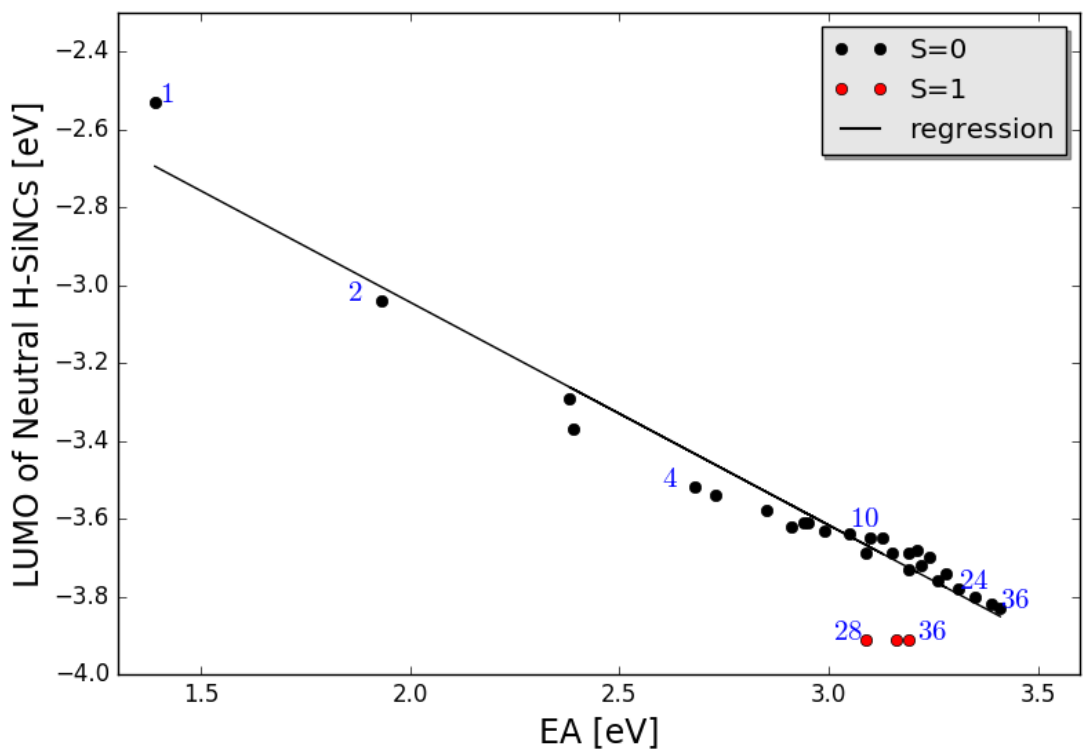

**Figure 6**. Correlation of the computed electron affinity (EA) for all H-SiNCs studied with the corresponding LUMO energy computed for the corresponding neutral H-SiNC. The numbers in the graph indicate the number of fused rings (N) for selected H-SiNCs. Here S represents the triplet and singlet states for S = 1 and S = 0, respectively.

In Figure 6, the calculated LUMO energy level versus electron affinity for all H-SiNCs are shown. The data shows two different electronic configurations based on the initial ground states of the neutral molecules. First, we observe that there is a correlation between the electron affinity (EA) with the DFT-computed LUMO energies of the neutral species ($E_{LUMO}$) with the corresponding initial spin state, having a linear regression of

$$E_{LUMO} = -0.573 * EA - 1.89 \ eV \quad (5)$$

Where both $E_{LUMO}$ and EA are in eV. The average absolute error in this relationship is 0.052 eV. The trend shows that LUMO energy levels generally decreased with increased EA. In addition, it is noted that the EA increases as function of the number of fused rings (N). These calculations indicate that both LUMO and EA can be independent to some extent, for instance, with respect to the number of fused rings.

# Summary and Outlook

In this study, we have shown that low hole and electron reorganization energies of H-SiNCs can be obtained. The most straightforward way of reducing these reorganization energies was achieved by increasing the number of fused rings, N. This is due to the extended $\pi$ conjugation of H-SiNCs. It was further found that there exists an inverse dependence between the electron reorganization energy and the number of fused rings. Moreover, silicene nanoclusters are promising candidates as electron acceptors (*n*-type) because they have high electron affinities and low reorganization energies for electron transfer, which result from a large conjugated system. EA and $\lambda_-$ can be further optimized by extending conjugation either through increasing the size of the system or by picking a specific direction, i.e., zigzag or armchair. We can also conclude that the spin state makes a significant change to the reorganization energy for corresponding silicene nanoclusters with $n_a > n_z$. In addition, the percentage of nonbonding character could play a crucial role in the internal reorganization. Therefore, understanding the charge transport mechanism in terms of reorganization energies of these H-SiNCs is a key point for the design and stability of H-SiNCs electronic components.

Moreover, having explored the reduction of the reorganization energy through two methods, we suggest another strategy for reducing the reorganization energy: the application of constraints that prevents or minimizes the restructuring (distortions) along the vertical axes for the silicene nanoclusters. This strategy has precedence from organic systems such as fullerenes, which have very low reorganization energies because they are extremely rigid. Therefore, this concept could be explored for H-SiNCs along with the advantage of using atomically thin Van der Waals materials and their ability to form heterostructures. This exploits the fact that these heterostructures can form junctions that are composed of *p*- and *n*-type semiconductors, each one being one unit cell thick, which exhibit completely different charge transport characteristics than bulk heterojuctions [42]. Thus, we can create a *p-n* junction by using individually contacted layers of *p*-type tungsten diselenide [43] and the *n*-type silicene nanoclusters to create an atomically thin *p-n* junction. This type of heterostructure would need to encapsulate the silicene nanoclusters within a rigid matrix that could avoid structural rearrangements. In addition, these types of systems may lead to unique material platforms of novel, high-performance electronic and optoelectronic devices.

## AUTHOR INFORMATION

**Corresponding Author**

*To whom correspondence should be addressed. Email: tvan@mit.edu

**Author Contributions**

All authors have given approval to the final version of the manuscript.

**Notes**

The authors declare no competing financial interest.

¶ Deceased author.

**ACKNOWLEDGMENT**

R.P.-P. would like to dedicate this paper to the memory of his mentor and advisor: M.S.D. " I feel very honored to have worked with Millie, including a series of weekend meetings, where she discussed both my work, and my career development. Thank you Millie for your advice and support, we all miss you at MIT ". R.P.-P. is also grateful to R. Olivares-Amaya for discussions on reorganization energy in organic systems. M.S.D., J. K., and R.P.-P. acknowledge the King Abdullah University of Science and Technology for support under contract (OSR- 2015-CRG4-2634). H.L.R. and S.F. acknowledge financial support from CONACyT (Grant 251684). J.L.M.C. acknowledges start-up funds from Florida State University and the Energy and Material Initiative and facilities at the High Performance Material Institute.

**Supporting Information Available**: A detailed discussion of the spin contamination for the cations and anions is given. CASSCF calculations for the system are given, which present an initial triplet state. Tables of bond length alterations for H-SiNC(4,9) with S=1 and S=0.

ABBREVIATIONS

DFT, density functional theory; H-SiNCs, silicene nanoclusters with hydrogen passivated edges.

# References

[1]   F. Xia, D. B. Farmer, Y. M. Lin, and P. Avouris, Nano Lett. **10**, 715 (2010).

[2]   L. G. De Arco, Y. Zhang, C. W. Schlenker, K. Ryu, M. E. Thompson, and C. Zhou, ACSNano **4**, 2865 (2010).

[3]   S. S. Varghese, S. Lonkar, K. K. Singh, S. Swaminathan, and A. Abdala, Sensors Actuators, B Chem. **218**, 160 (2015).

[4]   R. Raccichini, A. Varzi, S. Passerini, and B. Scrosati, Nat. Mater. **14**, 271 (2015).

[5]   J. Qiao, X. Kong, Z.-X. Hu, F. Yang, and W. Ji, Nat. Commun. **5**, 4475 (2014).

[6]   RadisavljevicB, RadenovicA, BrivioJ, GiacomettiV, KisA, B. Radisavljevic, A. Radenovic, J. Brivio, V. Giacometti, and A. Kis, Nat Nano **6**, 147 (2011).

[7]   S. Cahangirov, M. Topsakal, E. Akturk, H. Sahin, and S. Ciraci, Phys. Rev. Lett. **102**, 236804 (2009).

[8]   R. Pablo-Pedro, H. Lopez-Rios, S. Fomine, and M. S. Dresselhaus, J. Phys. Chem. Lett. **8**, 615 (2017).

[9]   Y. Ding and J. Ni, Appl. Phys. Lett. **95**, 83113 (2009).

[10]  L. Tao, E. Cinquanta, D. Chiappe, C. Grazianetti, M. Fanciulli, M. Dubey, A. Molle, and D. Akinwande, Nat. Nanotechnol. **10**, 227 (2015).

[11]  Y. Wang, J. Zheng, Z. Ni, R. Fei, Q. Liu, R. Quhe, C. Xu, J. Zhou, Z. Gao, and J. Lu, Nano **7**, 1250037 (2012).

[12]  W. L. Wang, S. Meng, and E. Kaxiras, Nano Lett. **8**, 244 (2008).

[13]  L. C. Campos, V. R. Manfrinato, J. D. Sanchez-Yamagishi, J. Kong, and P. Jarillo-Herrero, Nano Lett. **9**, 2600 (2009).

[14]  J. H. Schön, C. Kloc, and B. Batlogg, Phys. Rev. Lett. **86**, 3843 (2001).

[15]  J. L. Bredas, J. P. Calbert, D. A. da Silva Filho, and J. Cornil, Proc. Natl. Acad. Sci. **99**, (2002).

[16] N. M. Oboyle, C. M. Campbell, and G. R. Hutchison, J. Phys. Chem. C **115**, 16200 (2011).

[17] G. R. Hutchison, M. A. Ratner, and T. J. Marks, J. Am. Chem. Soc. **127**, 16866 (2005).

[18] G. R. Hutchison, M. A. Ratner, and T. J. Marks, J. Am. Chem. Soc. **127**, 2339 (2005).

[19] a Devos and M. Lannoo, Phys. Rev. B - Condens. Matter **58**, 8236 (1998).

[20] J.-L. J.-L. Brédas, J.-L. J.-L. Brédas, D. Beljonne, D. Beljonne, V. Coropceanu, V. Coropceanu, J. Cornil, and J. Cornil, Chem. Rev. **104**, 4971 (2004).

[21] J. C. Sancho-García and a J. Pérez-Jiménez, Phys. Chem. Chem. Phys. **11**, 2741 (2009).

[22] S. Larsson, A. Klimkans, L. Rodriguez-Monge, and G. Duskesas, Theochem-Journal Mol. Struct. **425**, 155 (1998).

[23] Y.-P. Sun, P. Wang, and N. B. Hamilton, J. Am. Chem. SOC **115**, 6378 (1993).

[24] V. Coropceanu, M. Malagoli, D. a da Silva Filho, N. E. Gruhn, T. G. Bill, and J. L. Brédas, Phys. Rev. Lett. **89**, 275503 (2002).

[25] M. Xu, T. Liang, M. Shi, and H. Chen, Chem. Rev. **113**, 3766 (2013).

[26] Q. H. Wang, K. Kalantar-Zadeh, A. Kis, J. N. Coleman, and M. S. Strano, Nat. Nanotechnol. **7**, 699 (2012).

[27] X. Ling, H. Wang, S. Huang, F. Xia, and M. S. Dresselhaus, Proc. Natl. Acad. Sci. **112**, 4523 (2015).

[28] A. N. Rudenko, S. Brener, and M. I. Katsnelson, Phys. Rev. Lett. **116**, 246401 (2016).

[29] S. Grimme, J. Antony, S. Ehrlich, and H. Krieg, J. Chem. Phys. **132**, 154104 (2010).

[30] L. Noodleman, J. Chem. Phys. **74**, 5737 (1981).

[31] (n.d.).

[32] D. J. Frisch, M. J.; Trucks, G. W.; Schlegel, H. B.; Scuseria, G. E.; Robb, M. A.; Cheeseman, J. R.; Scalmani, G.; Barone, V.; Mennucci, B.; Petersson, G. A.; Nakatsuji, H.; Caricato, M.; Li, X.; Hratchian, H. P.; Izmaylov, A. F.; Bloino, J.; Zheng, G.; Sonnenb, (2013).

[33] Y. C. Chang and I. Chao, J. Phys. Chem. Lett. **1**, 116 (2010).

[34] N. E. Gruhn, D. A. Da Silva Filho, T. G. Bill, M. Malagoli, V. Coropceanu, A. Kahn, and J. L. Brédas, J. Am. Chem. Soc. **124**, 7918 (2002).

[35] X. Amashukeli, J. R. Winkler, H. B. Gray, N. E. Gruhn, and D. L. Lichtenberger, J. Phys. Chem. A **106**, 7593 (2002).

[36] G. S. T. Jonathan C. Rienstra-Kiracofe and Henry F. Schaefer, Chem. Rev. **102**, 231

(2002).

[37] M. Y. Kuo, H. Y. Chen, and I. Chao, Chem. - A Eur. J. **13**, 4750 (2007).

[38] Y.-C. Chang, M.-Y. Kuo, C.-P. Chen, H.-F. Lu, and I. Chao, J. Phys. Chem. C **114**, 11595 (2010).

[39] C. R. Newman, C. D. Frisbie, A. Demetrio, S. Filho, and J. Bre, Chem. Mater. **16**, 4436 (2004).

[40] A. Facchetti, M. H. Yoon, C. L. Stern, H. E. Katz, and T. J. Marks, Angew. Chemie - Int. Ed. **42**, 3900 (2003).

[41] S. S. Zade and M. Bendikov, Chem. - A Eur. J. **14**, 6734 (2008).

[42] C.-H. Lee, G. Lee, A. M. van der Zande, W. Chen, Y. Li, M. Han, X. Cui, G. Arefe, C. Nuckolls, T. F. Heinz, J. Guo, J. Hone, and P. Kim, Nat. Nanotechnol. **9**, 676 (2014).

[43] X. Yu, M. S. Prévot, N. Guijarro, and K. Sivula, Nat. Commun. **6**, 7596 (2015).